\documentclass[namedreferences]{SolarPhysics}
\usepackage{epsfig}

\newcommand{\be}{\begin{equation}}
\newcommand{\ee}{\end{equation}}
\newcommand{\beq}{\begin{eqnarray}}
\newcommand{\eeq}{\end{eqnarray}}
\usepackage[all]{xy}
\newcommand{\ud}{\mathrm{d}}

\begin{document}
\begin{article}
\begin{opening}
\title{Enhanced Joule Heating in Umbral Dots}

\author{Chandan \surname{Joshi},
        Lokesh  \surname{Bharti} and
        S.N.A.  \surname{Jaaffrey}}
      
\runningauthor{JOSHI ET AL.}
\runningtitle{Enhanced Joule Heating in Umbral Dots}

\institute{Astronomy and Astrophysics Laboratory, Department of Physics, University College of Science, Mohanlal
                  Sukhadia University, Udaipur, 313001, India
                  \email{chandan-joshi@hotmail.com}\\ }

\date{Received ; accepted }

\begin{abstract}

We present a study of magnetic profiles of umbral dots (UDs) and its consequences on the Joule heating 
mechanisms. Hamedivafa (2003) studied Joule heating using vertical
component of magnetic field. In this paper UDs magnetic profile has been investigated 
including the new
azimuthal component of magnetic field which might explain the relatively larger enhancement of 
Joule heating causing more brightness near circumference of UD.

\end{abstract}
\end{opening}

\section{Introduction}
\label{Introduction}

Umbral dots (UDs) are bright features observed in the sunspot umbrae. Detailed study of such bright 
features play a key role in understanding the energy transport in sunspots and they exhibit most 
important physical parameters such as temperature, brightness, lifetime, magnetic field, size, 
mass outflows etc. For recent reviews on    the subject, see Solanki (2003), 
Thomas and Weiss (2004). 

  There are two models for the UDs. First suggested by Parker (1979a, b) and then 
subsequently by Choudhuri (1986). Parker suggested in his sunspot model that the region below 
the visible surface termed as positive Wilson depression is made up of individual flaring 
flux tube embedded in the field free plasma. These tubes merge into an apex like single 
flux tube just above the umbral surface (negative Wilson depression level) of the sunspot. 
The expanding and rising up-flow of plasma creates a gap and develops UDs when it reaches 
to zero Wilson depression surface in sunspot. Further Choudhuri (1986) also showed that if 
the pressure of the 
plasma plume
increases at the apex of the static configuration of a field free gas column, then it rises to
certain height where the magnetic field pressure suddenly becomes negligible. Finally the
trapped field free gas bursts in to a column at a speed of about 10 km~s$^{-1}$ forming a UD.
In another magnetohydrodynamic model, UDs are considered as tops of convective cells 
present in homogeneous magnetic field (Knobloch and Weiss, 1984). Later Degenhardt and Lites (1993) 
speculated that UDs are thin vertical magnetic tubes with a reduced
magnetic field strength, a temperature enhancement and material upward flow from the
bottom, embedded in a stationary sunspot umbra.

Some experimental studies have revealed that the lifetime of UDs is in the
range from 15 min to 2 hrs with average diameter of 150--300 km and the relative 
intensities with regard to umbra of sunspot vary from 1.1--2.6 (Lites {\it et al.}, 1991; Ewell,
1992; Sobotka {\it et al.}, 1997a; Tritschler and Schmidt, 2002; Sobotka and Hanslmeier, 2005).
The number of investigators have also observed mass upward motion within the range of 0.3--1 km~s$^{-1}$
(Pahlke and Wiehr, 1990; Lites {\it et al.}, 1991; Wiehr, 1994; Rimmele, 1997, 2004).

  The role of magnetic profile has been unique for some specific properties 
of UDs and in one of the studies of magnetic profiles, Choudhuri (1986) established that there 
has been a reduced magnetic strength in the UD column relative to the surrounding umbra.
Some spectroscopic observations of Adjabshirzadeh and Koutchmy (1983), Pahlke and Wiehr
(1990), and Wiehr and Degenhardt (1993) have shown that positive magnetic field gradients in
umbral dots are of about 20\%, whereas in the high resolution study of Lites {\it et al.} (1991) have 
revealed that magnetic 
strength is not significantly different from that the surrounding darker portions of the 
umbrae. Also recent study (Socas-Navarro {\it et al.}, 2004) showed that 
there is more inclined fields in UDs and the vertical gradient of field may have opposite signs 
in the UD and dark background. Since all observations are obtained often with a low spatial
 resolution restricted for a small area so that the
complete magnetic vector remains unknown. Thus these observational results point out
that a good knowledge of magnetic vector field is crucially required.
However, in recent simulation study of Sch\"ussler and  V\"ogler
(2006),they have manifested that nearly field free upflow plumes and the UDs are a natural result
of convection in a strong initially monolithic magnetic field.  

In the subsequent Section 2 magnetic field
profile for UDs has been discussed whereas in Subsection 2.2, we have tried to
develop the modified magnetic field profile by introducing a proposed azimuthal
component $B_{\phi}(r)$ crucially required with vertical component $B_{z}(r)$
and its effect on Joule heating. The resultant magnetic vector obtained for the UDs might 
be attributed to the increased current density which probably may enhance Joule heating power 
causing a relatively more brightness near the circumference as compared to the center of the UDs.

\section{Magnetic Profile of Umbral Dots and Joule Heating}
\label{magnetic field}

We still know very little about the structure and the nature of the magnetic field of UD, as there are few 
high-resolution evidences available (Lites {\it et al.}, 1991; Socas-Navarro {\it et al.}, 2004) for the vector 
magnetic field that can provide correct magnetic field structure in UDs. In spite of these 
observational challenges, the right choice of magnetic profile may produce better consistency 
with observations. In the next Subsection 2.1 we first give an overview of vertical magnetic field 
 and then we propose an additional azimuthal magnetic field with the vertical 
component in the subsequent Subsection 2.2.

\subsection{Vertical component of magnetic field in umbral dots}
\label{vertical magnetic field}

As a matter of fact Joule heating power is partially responsible for brightness of the UDs and is
directly governed by current density, which in principle is attributed to magnetic field
profile (Garcia de la Rosa, 1987). Thus it was assumed for simplicity that magnetic field
vector has only vertical parallel component in the UD column (Hamedivafa, 2003)
and is a function of distance from axis of the UD column as

\begin{equation}
\mathbf B =  B_{z}(r) \mathbf a_{z}
\end{equation}       
where $\hat a_{z}$ is the unit vector along z-axis normal to the photosphere of the Sun and 
{\it r} is the
radial component in cylindrical coordinates $(r,\phi,z)$. Hamedivafa (2003) investigated Joule 
heating as brightening mechanism for umbral dots and he assumed one of the magnetic profiles given as
\begin{equation}
 B_{z}(r) = B_{0}- \gamma B_{0} \exp {\frac{ -r^2}{u^2}}  
\end{equation}
where $B_{0}$ is saturated magnetic field in umbra of the sunspot, $\gamma$ is fraction of the 
field redution and {\it u} is the maximum radius of the UD column at any instant. 
Hamedivafa and Sobotka (2004) found direct 
observational evidence for Joule heating in some of the UDs. 

\subsection{Azimuthal component of magnetic field in UDs}

A prudential overview of the existing models for the magnetic field of UDs is
required since the magnetic nature of bright features in sunspot umbrae is not fully
understood yet. The observational results of Lites {\it et al.} (1991) pointed out that UDs magnetic field 
strength is not very different from umbral field and also Socas-Navarro {\it et al.} (2004) showed that UDs 
have inclined fields. This promulgated the following two major assumptions which have been
contemplated in theoretical investigations:

\begin{enumerate}
\item $B_{z}(r)$ is the force free axial vector field for the expanding and rising 
parallel axial flow of plasma plume in UD column. It develops cusp-like shape due to
steeper pressure gradient of piled-up plasma in UDs leading to a drastic decrease
of magnetic field strength in the upper layer of plumes than the proximate circumference
of UD.
\item Steady flow of material in UD column does not give a temporally changing electric
field i.e. there is no changing electric flux.
\end {enumerate}

These assumptions deduce a basic magneto-hydrodynamic equation

\begin{equation}
(\nabla \times {\mathbf B})_{z} \propto \mu \ J_{z}
\end{equation}    
which is valid for azimuthal component. Where  $\mu$ is the magnetic permeability across
the active region and the current density $J_{z}$ is associated with a convectively 
unstable and oscillating vertical slab of plasma, sandwiched between the regions of 
$B_{z}(r)$ with vertical wavelength of 100 km and of period 100 s (Parker, 1979b). 
Moreover it may be expressed as a possible electric current parallel to the force free 
$B_{z}(r)$ within $r\le u$ column as
\begin{equation}
 I =\oint  J_{z} \cdot da
\end{equation}    
Here {\it r} is the radial distance from the UD axis; {\it u} is the maximum radius of UD at which
an undisturbed magnetic field strength of umbra exists. The high temperature of 
the hot unstable outflowing material would be sufficient to yield an ionized form, creating 
enough electrical current in these column. These concepts led us to
believe that there should be azimuthal magnetic component $B_{\phi}(r)$ attached to the thin
UD column, due to electric current of material at high
temperature. Therefore, the magneto-hydrodynamic calculations under some boundary conditions
for evolution of these flux tubes, diagnose UDs of enhanced temperature and high
intensity relative to surrounding umbra. However this current would be able to
generate azimuthal magnetic field $B_{\phi}(r)$ with the help of Ampere  circuital law, Equation (3),
and $B_{\phi}(r)$ may be expressed as
\begin{equation}
B_{\phi}(r) = \frac{\gamma \mu I r}{2 \pi u^{2}}  \qquad \textrm{when}~~~ r < u
\end{equation}    
\begin{equation}
B_{\phi}(r) = \frac{\gamma \mu I }{2 \pi r}  \qquad \textrm{when}~~~ r > u.
\end{equation} 
Here $\gamma$ is the fraction of the field strength reduction on the central axis of the UD column
with range $1\geq\gamma\geq0$. Hamedivafa (2003) has revealed that the magnetic field reduction is 
not beyond the radius of the bright UDs but saturates after $r/u\approx2$. The contribution of $B_{\phi}(r)$ is linear from the
center to the circumference of the UD and then hyperbolically decreases beyond the UD
column. The inclusion of $B_{\phi}(r)$ in the proposed model may be 
justified on the basis of following assumptions:
\begin{enumerate}
\item The center of the azimuthal component coincides with the symmetric axis of the UDs. It
has been assumed to resolve the ambiguity between center of $B_{\phi}(r)$ and vertical
component $B_{z}(r)$ which yields more or less fine radial
magnetic structure inside UDs producing a temperature stratification and a height
independent values for magnetic field strength with respect to line of sight.
\item Magnetic field of the UDs consists of the two components, $B_{\phi}(r)$ and $B_{z}(r)$ 
to provide a
consistent feature of magnetic profile in such a way that the magnetic field inside an UD is weaker than
surrounding field and large magnetic field gradient is present at the boundary of UD. The component 
$B_{\phi}(r)$ might contribute more power to Joule heating. 
\item $B_{\phi}(r)$ at the circunference of an UD becomes comparable to the observable magnetic field, $B_{0}$ of umbrae.
\end{enumerate}

Thus resultant $\mathbf B(r)$ may be written as 
\begin{equation}
\ {\mathbf B(r)} = B_{\phi}(r) {\mathbf a_{\phi}} + B_{z}(r) {\mathbf a_{z}}.
\end{equation}

\begin{figure}    
\begin{center}
\psfig{file=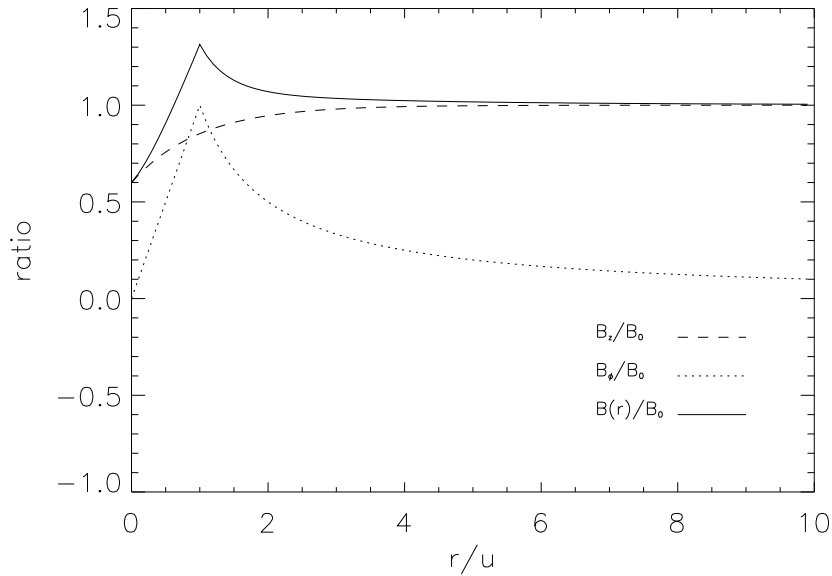,width=12.cm,clip=}
\vspace*{0.1cm}
\hspace*{1.5cm}
\caption{Normalized magnetic field strength of UD versus normalized distance from the UD centre.}
\label{magneticprofile}
\end{center}  
\end{figure}

Normalised $ B(r)$, $ B_{\phi}(r)$ and $ B_{z}(r)$ with $B_{0}$ are plotted as shown in Figure 1.
The effective current density in UDs is calculated by the magnetic field as
\begin{equation}
\ \mathbf J_{\rm T} = \frac{c}{4 \pi} \nabla \times \mathbf B(r) 
\end{equation}

or the value of $\bf J_{\rm T}$ is calculated in cylindrical co-ordinate system 
by the followig determinant
\begin{displaymath}
\vec J_{\rm T} =
\left| \begin{array}{ccc}
\mathbf a_{r} & \mathbf a_{\phi} & \mathbf a_{z} \\
\frac{\ud}{\ud r} & \frac{\ud}{r \ud \phi} & \frac{\ud}{\ud z} \\ 
0 & \frac{\gamma \mu Ir }{2 \pi u^2}   & B_{0}- \gamma B_{0} \exp {\frac{ -r^2}{u^2(t)}}
\end{array} \right|.
\end{displaymath}

\begin{equation}
\ \mathbf J_{\rm T} = \left [ \left ( \frac{-c}{4 \pi} \gamma B_{0} \exp{ \frac{-r^2}{u^2} } \right) \times \frac{2r}{u^2} \right ] \mathbf a_{\phi}+\left[\frac{c \gamma \mu I}{8 \pi^2 u^2} \right] \mathbf a_{z}
\end{equation}

Now the Joule heating power can be given as:
\begin{equation}
\ Q^{\prime} = \frac{4 \pi}{c^2} \int_{0}^{\infty} \eta J_{\rm T}^2 (2 \pi r) dr,
\end{equation}
where $\eta$ is electrical resistivity.
\begin{equation}
 J_{\rm T}^2 = \mathbf J_{\rm T} \cdot \mathbf J_{\rm T} = \frac {c^2}{16 \pi^2} \gamma^2 B_{0}^2 \frac{4r^2}{u^4} \exp{\left ( \frac{-2r^2}{u^2} \right )} + \frac{c^2 \gamma ^2 \mu ^2 I^2}{64 \pi^4 u^4}. 
\end{equation}
From Equation (10) and (11), we get
\begin{equation}
\ Q^{\prime} = \frac{2 \eta \gamma^2 B_{0}^2 } {u^4}  \left ( \frac{u^4}{8} \right) + \frac{\eta \gamma^2 \mu^2 I^2 u^2}{4 \left ( 4\pi^2 u^2 \right )}.
\end{equation}
We assume that when $r \to u$ then $ \vert B_{\phi}(r) \vert \to \vert B_{0}(r) \vert $ , i.e. the maximum strength of azimuthal component 
at the boundary of UD is comparable with the undisturbed magnetic field $B_{0}$
of the umbra outside of the UD. Thus from Equation (5) we get

\begin{equation}
 B_{0} = \frac {\gamma \mu I u}{2 \pi u^2} = \frac {\gamma \mu I}{2 \pi u} 
\end{equation}
or 
\begin{equation}
 \mu^2 I^2 = \frac {B_{0}^2 \left ( 2 \pi u \right ) ^2 }{\gamma^2}. 
\end{equation}
Let
\begin{equation}
 U_{\gamma B} = \frac{\gamma^2 B_{0}^2}{ 8 \pi}. 
\end{equation}
With the help of Equations (12) -- (15), the final power can be given as
\begin{equation}
\ Q^{\prime} = \frac{4 \pi \eta U_{\gamma B}} {2}  \left [1 + \frac{1}{\gamma^2} \right]. 
\end{equation}
The Joule Heating power is proportional to $2 \pi \eta U_{\gamma B}$ whereas $ Q= 2 \pi \eta U_{\gamma B}$ is calculated by 
Hamedivafa (2003). The factor in the square bracket represents the enhancement in the
power due to the current generated from the magnetic field at $r$ close to $u$. For
a smaller value of $ \gamma $, larger would be the Joule heating power. This justifies the special
features observed in the experimental studies of Lites {\it et al}. (1991) that the field strength
within the umbrae vary at large scale (1400 -- 2400 Gauss) and this large scale variation of field being
inversely correlated with the umbral dot intensity.

\section{Discussion and Conclusions}
\label{conclusions}

The result presented in this report depends upon the following crucial assumptions:
\begin{enumerate}
\item The $B_{\phi}(r) $ is developed just at the middle of UDs and increases linearly with 
$r$ inside the UD diameter up to the maximum value $B_{0}$. It decreases hyperbolically outside.
\item The current density $ \bf J_{\rm T}$ is uniform within the UDs.
\item Magnetic profile of UDs consists of at least two magnetic components, $B_{\phi}(r) $ and $B_{z}(r) $
which stem not necessarily from the same geometrical height.
\end{enumerate}

We conclude enhanced field on peripheral surface of UD as shown in Figure 1. The large peripheral 
flux is attributed to the
additional azimuthal magnetic field component, which is produced by axial current {\it I} due to 
intrusion of hot plasma. Thus modified UD model obtained by adding 
$B_{\phi}(r) $ into $B_{z}(r) $ helps to explain variation in brightness 
of the UDs. The relative Joule heating
\begin{equation}
\ \vartriangle Q = \frac {Q^{\prime}}{Q} = 1+ \frac{1}{\gamma^2} 
\end{equation}

\begin{table*}[h]
\caption[r]{ The relative Joule heating with diffrent values of $\gamma$ }
\label{tab1}
\begin{tabular}{rrrrrrrrrrrr}
\hline 
$\gamma$ & 0.1 & 0.2 & 0.3 & 0.4 & 0.5 & 0.6 & 0.7 & 0.8 & 0.9 & 1.0\\
$ \vartriangle Q $ & 101.0 & 26.0 & 12.11 & 7.25 & 5.00 & 3.78 & 3.04 & 2.56 & 2.23 & 2.00\\
\hline
\end{tabular}
\end{table*}

Fractional Joule heating is much more at lower values of $ \gamma $ but at higher values
decreases slowly. The estimated Joule heating power is expected to justify the inverse 
correlation of brightness and magnetic field gradient (Lites {\it et at.}, 1991) and 
the resultant magnetic vector supports more inclined magnetic field vector 
(Socas-Navarro {\it et al.}, 2004) of the UD. 

Moreover, these conclusions can be justified by high resolution spectropolarimetric data from 
ground based instrument such as Diffraction Limited Spectropolarimeter (\textbf {DLSP})
(Sankarsubramanian {\it et al.}, 2004) at Dunn Solar telescope (NSO) and the
spectropolarimeter onboard the satellite Hinode.


\acknowledgements
    The authors are thankful to anonymous referee for constructive comments
to improve the presentation of this manuscript and Prof. Takashi Sakurai 
for helpful editorial work.
C.J. acknowledges Shrinathji Institute of Technology and Engineering
for encouragement. L.B. is thankful to Education Department Govt. of Rajasthan 
and Bal Shiksha Sadan Samiti, Udaipur for encouragement.

\end{article} 
\end{document}